\documentstyle[preprint,aps,epsfig,12pt]{revtex}
\begin{document} 
%\input{first_page}
%for journal submition use instead the following format
%\documentstyle[preprint,aps,epsfig,12pt]{revtex}
\begin{flushright}CEBAF-TH-97-43\end{flushright}\vspace{2.0 cm}
\begin{center} 
The  Decays $K_L\rightarrow \ell^+\ell^-\ell'^+\ell'^-$ Revisited
\begin{large}
\end{large}

\vskip 0.5in

Longzhe Zhang and J. L. Goity

\vskip 0.2in

{\it Department of Physics, Hampton University, Hampton, VA 23668, USA
\\
and \\
  Jefferson Lab,  12000 Jefferson Avenue, Newport News, VA 23606, USA. }

\end{center}

\thispagestyle{empty}

\begin{titlepage}
\thispagestyle{empty}
\title{ The  Decays $K_L\rightarrow \ell^+\ell^-\ell'^+\ell'^-$
Revisited}
%\vspace*{1.cm}\\
\author{ Longzhe Zhang and J. L. Goity}
\address{
Department of Physics, Hampton University, Hampton, VA 23668, USA \\
and \\
  Jefferson Lab,  12000 Jefferson Avenue, Newport News, VA 23606, USA.} 
\maketitle  
\begin{abstract}
 The double lepton pair decay modes of the $K_L$ meson are analyzed
including all
  contributions of order $p^6$ in Chiral Perturbation Theory.
  The experimentally established $e^+ e^- e^+ e^- $  mode and the
recently 
  observed $e^+ e^- \mu^+ \mu^-$ mode are discussed in detail.
\end{abstract}
\vspace{20mm}
\pacs{{\tt$\backslash$\string pacs\{12.15.-y, 12.39.Fe, 13.25.Es \}}}
\end{titlepage}
\newpage
In this note we reconsider the double lepton pair  decays 
 $K_L \rightarrow \ell^+\ell^-\ell'^+\ell'^-$ that
 were studied many years ago by Miyazaki and Takasugi$\cite{Miyazaki}$.
The two    
 modes of  main interest 
 are   the $K_L \rightarrow e^+e^-e^+e^-$ and the 
 $K_L\rightarrow e^+e^-\mu^+\mu^-$ decays. The first proceeds 
 at a rate well determined experimentally$\cite{PDG}$  with a  
branching ratio equal 
 to $4.1\pm 0.8\times 10^{-8}$, and its 
   interest resides partly in the possibility of determining the form 
   factor associated with the virtual $\gamma$
    couplings to the $K_L$ as well as the interesting interference  
    effect derived from the  identical leptons in the 
    final state. Both issues are analyzed here, and the main conclusions
are 
     that the form factor effect is small
(4\%), requiring a substantial experimental improvement over the current
error 
of about 20\%, and the interference
 effect is tiny (0.5\%) and most likely beyond experimental access.  
 The second decay has   been recently observed
  for the first time  by the E799 collaboration at Fermilab
$\cite{E799}$.
   This experiment quotes a branching ratio 
  of a few parts per billion, a remarkable improvement over the
previous  upper 
  bound  of $4.9\times 10^{-6}$ $\cite{PDG}$.
    We discuss in this case the effects of a non-trivial form factor,
that according
     to our result produces an increase of the 
     branching ratio by 30\%, which could therefore be tested in the
future as the
      experimental accuracy in this mode improves.

   In the following we neglect a contribution to the amplitude due to CP
violation.
    This contribution is of order $p^4$ but
    suppressed by the  CP violating mixing parameter $\epsilon$ which is
of order
     $10^{-3}$  (${\rm Re}(\epsilon)=1.6\times 10^{-3}$,
$\phi_{\epsilon}\simeq 43.5 ^o$). 
     Thus, the CP violating contributions to the widths considered here
are  suppressed by a 
     factor approximately equal to 
     $\mid \epsilon \mid  \Lambda_{\chi}^2/M_K^2 \sim 1 \%$, where
     $ \Lambda_{\chi} =4\pi \, F_\pi$. 
      Thus, in the limit of CP conservation, the
$K_L\rightarrow\gamma^*\gamma^*$ 
      amplitude has the most general form:
   \begin{eqnarray}
 A(K_L\rightarrow\gamma^*\gamma^{\prime *})&=&{\cal {F}}(t,t^\prime)
\;\epsilon^{\mu\nu\rho\sigma} \;\epsilon_{\mu} k_{\nu}
 \epsilon^{\prime}_{\rho} k^{\prime }_{\sigma}.
\end{eqnarray}
Here $\gamma^*$  and $\gamma^{\prime *}$  are virtual photons with
respective 
invariant mass squared 
$t=k^{2}$
 and $ t^\prime=k^{\prime 2}$. ${\cal {F}}(t,t^\prime )$ is the form
factor of order $p^6$
  in the chiral expansion 
 that  has a $t$ and $t'$ independent contribution   $F_1$, due to the
$\pi^0$, $\eta$ and 
 $\eta '$ poles$\cite{Donoghue}$, plus a $t$ and $t'$ 
 dependent contribution $F_2(t,\,t')$ from one chiral loop plus
counterterms.   $F_1$ is 
 entirely fixed up to its sign by 
 the $K_L\rightarrow\gamma\gamma$ decay, and $F_2(t,\,t')$ is given by
$\cite{Zhang1}$:
\begin{equation}
 F_{2} (t,t^\prime) =\frac { \alpha_{\rm em} C_{8} } { 192 \pi ^{3}
F_{\pi} ^{3} } 
  \, \{-(a_{2} + 2 a_{4} )\; D (t,t^\prime,\mu) +C(\mu)\; (t+t')\} ~~ ,
\end{equation}
where the  counterterm has eliminated an UV divergence proportional to
$(t+t')$,  
$F_\pi=93$ MeV is the pion decay constant, and 
\begin{eqnarray}
    D(t,t^\prime,\mu)&=&(t+t^\prime)\; [\frac{10}{3} - 
     ( \log{\frac { M_{K} ^{2} } { \mu ^{2} }} + 
     \log{\frac { M_{\pi} ^{2} } { \mu ^{2} } })]     \nonumber \\ 
 &+ & 4 \left[ F( M_{\pi}^{2},   t ) + F( M_{K}^{2},  t)+  F(
M_{\pi}^{2},  t^\prime ) 
 + F( M_{K}^{2},   t^\prime)\right]  ~~ ,
 \end{eqnarray} 
with the chiral logarithms  contained in the function $F( m ^{2}, ~t)$:
\begin{eqnarray}
F( m ^{2}, ~t) &\equiv & \left(( 1 - \frac {y} {4} )\; \sqrt { \frac {y
- 4} {y} } 
\log{\frac { \sqrt {y} + \sqrt {y  - 4 } }
 { - \sqrt {y } + \sqrt {y - 4 } }} - 2 \right) m^{2} ,\nonumber \\ y  
 &\equiv & \frac{t}{ m^{2}}.
\end{eqnarray}
The coefficient  $C(\mu)$ in the counterterm is determined by the fit to
the Dalitz decays  
as discussed 
in  $\cite{Zhang1}$.
 There are two scenarios distinguished by the sign of $F_1$,  where the
relative size of 
 the counterterm 
 to the chiral logarithms is different. 
 In one scenario the counterterm (defined at the subtraction scale
$\mu=M_\rho$)
  provides more than 90\% 
 of the contribution to the form factor's slope,
  while in the other that fraction is reduced to about 75\%.

  Since ${\cal{F}}(0,t)={\cal{F}}(t,0)=F(t)$, where $F(t)$ is the form
factor of the 
  Dalitz decays studied in $\cite{Zhang1}$, 
  we use the results obtained in that reference by fitting the   data
$\cite{Dalitz}$:
\begin{eqnarray}
{\rm Scenario} ~~1:&&\nonumber\\
F_1&=& 0.89\;\frac{\alpha_{\rm em} C_8}{2 \pi F_\pi},
~~~  a_2+2 a_4= -0.3\pm 0.3 ,~~~  C(\mu=M_\rho)= 14.2\pm 7.3 
\nonumber\\
{\rm Scenario} ~~2:&&\nonumber\\
F_1&=&-0.89\;\frac{\alpha_{\rm em} C_8}{2 \pi F_\pi},~~~  a_2+2
a_4=1.5\pm 0.3,~~~  
 C(\mu=M_\rho)= -10.3\pm 7.3 \end{eqnarray}
Here, $C_8=3.12\times 10^{-7}$ is the octet coupling in the non-leptonic
weak
 interaction effective Lagrangian of order $p^2$.

The decay amplitude has one piece if the final lepton pairs are
different, and
 two pieces if they are identical.
 In this latter case the two amplitudes are: 

\begin{eqnarray}
A_1 &=& e^2 F(t,t^\prime )\;\epsilon^{\mu\nu\rho\sigma} \;
\frac {(p_{+}+ p_{-})_{ \nu} (p_{+} ^{\prime} 
+ p_{-} ^{\prime} )_{\sigma} }{ t t^{\prime} } \nonumber \\ &
 \times & \bar{u} ( p_{-} ) \; \gamma _{\mu} \; 
 \upsilon ( p _{+} ) \;\bar{u} (p _{-} ^\prime ) \;
 \gamma _{ \rho } \; \upsilon ( p ^\prime _{+} ) \; .
\end{eqnarray}

\begin{eqnarray}
A_2 &=& - \; e^2 F(s,s^\prime )\;\epsilon^{\mu\nu\rho\sigma} 
\;\frac {(p_{+} + p_{-} ^\prime )_{ \nu} (p_{+} ^{\prime} 
+ p_{-} )_{\sigma} } { s s^{\prime} } \nonumber \\ & \times & 
\bar{u} ( p ^\prime _{-} ) \;
\gamma _{\mu} \; \upsilon ( p _{+} ) \; 
\bar{u} (p _{-} )\; \gamma _{\rho}\; \upsilon ( p ^\prime _{+} ) \; ,  
\end{eqnarray}
 where $p_+$ is the momentum of $\ell^+$, etc., and
\begin{eqnarray}
t &=& (p_{+}  + p_{-} )^2 \; ,~~~~
t ^\prime = (p_{+} ^{\prime} + p_{-} ^{\prime} )^2 \; , \nonumber \\
s &=& (p_{+}  + p_{-} ^ \prime  )^2 \; ,~~~~
s ^\prime = (p_{+} ^{\prime} + p_{-}  )^2 \; .
\end{eqnarray}
In the case of distinguishable lepton pairs only one appears, say $A_1$. 
The decay width is obtained by 
summing over the lepton  spins and integrating   over the four particle
phase space.
 We   checked the results obtained  
in $\cite{Miyazaki}$  and refer to it for further details.

 In the $K_L\rightarrow \ell^+ \ell^- \ell^+ \ell^- $ decay we have:
 \begin{equation}
\Gamma(K_L\rightarrow \ell^+ \ell^- \ell^+ \ell^- ) =
\frac{1}{2}\; \Gamma _1(K_L\rightarrow \ell^+ \ell^- \ell^+ \ell^- )+
\frac{1}{2}\; \Gamma _2 (K_L\rightarrow \ell^+ \ell^- \ell^+ \ell^- )
\end{equation}
with
\begin{eqnarray}
\Gamma _{1} &=& \int \; \sum_{\rm spins} \mid A_{1~{\rm or}~2} \mid ^2
\; d \Phi \; ,
 \nonumber \\
\Gamma _{2} &=& \int \;  {\rm Re}( \sum_{\rm spins}  A_1 A_2 ^* )\; d
\Phi \; ,
\end{eqnarray} 
where $d \Phi$ represents the four body phase space volume element. 
 On the other hand, in the  $K_L\rightarrow e^+ e^- \mu^+ \mu^- $  decay
we have instead:
\begin{equation}
\Gamma(K_L\rightarrow e^+ e^- \mu^+ \mu^-  ) =
  \Gamma _1(K_L\rightarrow e^+ e^- \mu^+ \mu^-   )
  \end{equation}

For convenience we refer to the $K_L\rightarrow \gamma \gamma$ width.
Defining $\rho \equiv \Gamma / \Gamma ( K_L \rightarrow \gamma \gamma )$
and  $\rho_{\rm interference} \equiv \frac{1}{2}\;\Gamma_2 / 
\Gamma( K_L \rightarrow \gamma \gamma )$,
we obtain the results shown in the table:

\newpage

\vspace{10mm}
\begin{center}
\begin{tabular}{|c|c|c|c|c|}  \hline  ~ & ~ & ~ & ~\\

~~~Decay mode~~~ & ~~~Analysis~~~    &  $  \rho _{\rm interference}  $ 
&  $  \rho  $
 \\ & & & \\ \hline   & & &\\
 $e^{+}e^{-}e^{+}e^{-}$ & Ref \cite{Miyazaki} & ~~~$-0.35\times 10^{-5}
$
 ~~~ & $5.89 \times 10^{-5} $ \\ 

 & No form factor   & 

 $ -0.036 \times 10^{-5}  $ &  $ 6.26 \times 10^{-5} $ \\
 & With form factor    & 

$ -0.048 \times 10^{-5} $ & $ 6.50 \times 10^{-5} $\\
&   & 

$ -0.047 \times 10^{-5} $  & $  6.48 \times 10^{-5}$  \\ & & & \\
\hline   & & &\\
$ e^{+} e^{-} \mu^{+} \mu^{-} $ & Ref \cite{Miyazaki}   &
  0 & $ 1.42\times 10^{-6}$\\
&  No form factor   & 

 0   &  $ 1.71 \times 10^{-6} $ \\
 &  With form factor & 

 $ 0 $  & $ (2.20 \pm 0.25)\times 10^{-6}$ \\ 

& &  

$ 0  $ & $ (2.18 \pm 0.25) \times 10^{-6}$ \\ & & & \\ \hline   & & &\\ 

 $  \mu^{+} \mu^{-} \mu^{+} \mu^{-} $ & Ref \cite{Miyazaki}    &
$   -0.051 \times 10^{-9} $  & $  0.946 \times 10^{-9} $ \\   

& No form factor &   

   $ -0.051 \times 10^{-9}$  &  $ 0.93 \times 10^{-9}  $\\
& ~~~ With Form Factor ~~~   & 

$ -0.077 \times 10^{-9} $ &  $ (1.30\pm 0.15) \times 10^{-9} $\\ 

& &  

$  -0.072  \times 10^{-9} $  &~~~ $ (1.35 \pm 0.15) \times 10^{-9}$ ~~~
\\  & & & \\ \hline  
\end{tabular}

\vspace{6mm}

\parbox{4.5in}{ { {\bf TABLE}: The results of Ref \cite{Miyazaki}
correspond to
a   point like form factor. The results of this work with a form factor
are given
 respectively for the two scenarios of equation (5).  }} 

 \end{center}

 \vspace{3mm}

In the $e^+ e^-e^+ e^-$ mode there is a small effect due to the form
factor 
 leading to an increase of the width by about 4\%.
 Given the current experimental error of almost 20\%, it seems that  a 
 test of the form factor can be achieved in the foreseeable future. 
  On the other hand, we find that the interference term due to the
identity of 
   particles is a factor ten smaller than that obtained
   before $\cite{Miyazaki}$, and it represents a correction of 0.5\%, 
   which seems beyond experimental access. Our prediction is
    consistent with the experimental rate: 
    $BR(K_L\rightarrow e^+ e^-e^+ e^-)|_{\rm Theory }=3.85\times
10^{-8}$, and 
    $BR(K_L\rightarrow e^+ e^-e^+ e^-)|_{\rm Exp. }=(4.1\pm 0.8)\times
10^{-8}$.
The relative size of the interference effect is  larger in the $\mu^+
\mu^-\mu^+ \mu^-$ mode,
 but alas,
 the total branching ratio for this decay is 
predicted to be about
$ 8 \times 10^{-13}$, and clearly outside the scope of future
experiments.
 Finally, the $e^+ e^- \mu^+ \mu^-$ mode shows a sizeable 
effect due to the form factor that leads to  an increase of the width by
about 30\%.
 The recent  first-time observation of this decay mode by the E799
collaboration at
  Fermilab $\cite{E799}$ furnishes a first experimental 
 determination of the branching ratio, namely  
 $BR(K_L\rightarrow e^+ e^-\mu^+ \mu^-)|_{\rm Exp.
}=2.9^{+6.7}_{-2.4}\times 10^{-9}$.
  Our result is 
  $BR(K_L\rightarrow e^+ e^-\mu^+ \mu^-)|_{\rm Theory }=(1.30\pm
0.15)\times 10^{-9}$, 
  consistent with the measured value. 
  There is here a strong promise that the reduced error bars resulting
from 
  the future experimental efforts will  permit
   to show the effect due to the non-trivial form factor. This is
clearly the
    most interesting mode for further experimental study.

As one would have expected, the analysis of the Dalitz decays is enough
to
pin down the predictions  for the double lepton pair decays, and the two
 scenarios resulting from that analysis give essentially the same
results.

\newpage

 \bibliographystyle{unsrt}

\begin{thebibliography}{99}
\bibitem{Miyazaki}  T. Miyazaki and E. Takasugi, Phys. Rev. D8, (1973)
2051.
\bibitem{PDG} R. M. Barnett et al., Review of Particle Properties,  
Phys. Rev. D54, 1 (1996), and 1997 off-year partial 
update URL:http:$//$pdg.lbl.gov. 
\bibitem{E799} P. Gu et al., Phys. Rev. Lett. 76, 4312 (1996). 
\bibitem{Donoghue} J. F. Donoghue, B. R. Holstein and Y.-C.R. Lin, Nucl.
Phys. B277, 651 (1986).\\
J. L. Goity,  Z. Phys. C34, (1987) 34.
\bibitem{Zhang1} J. L. Goity and Longzhe Zhang, Phys. Lett. B 398, 387
(1997).  
\bibitem{Dalitz} G. D. Barr et al., Phys. Lett. B 240, 283 (1990).\\
K. E. Ohl et al., Phys. Rev. Lett. 65, 1407 (1990).\\
J. Enagonio et al., Phys. Rev. Lett. 74, 3323 (1995).
%\bibitem{ } 

\end{thebibliography}

\end{document}